\documentclass[12pt,JASA]{article}
\usepackage{amsmath, amssymb}
\usepackage{graphicx}
\usepackage{natbib}
\usepackage{geometry}
\usepackage{setspace}
\usepackage{booktabs}
\usepackage{float}
\usepackage{caption}
\usepackage{color}
\usepackage{lineno}
\usepackage{psfrag,epsf}
\usepackage{enumerate}
\usepackage{enumitem}
\usepackage{bm}
\usepackage{url} 
\usepackage{authblk} 

\usepackage{amsthm}  
\newtheorem{theorem}{Theorem}[section]  

\newtheorem{assumption}{Assumption}

\theoremstyle{definition}

\theoremstyle{remark}

\newcommand{\blind}{0}

\def\spacingset#1{\renewcommand{\baselinestretch}%
{#1}\small\normalsize} \spacingset{1}

\title{Penalized Empirical Likelihood for Doubly Robust Causal Inference under Contamination in High Dimensions}
\author[1]{Byeonghee Lee}
\author[2]{Sangwook Kang}
\author[3]{Ju-Hyun Park}
\author[4]{Saebom Jeon}
\author[5]{Joonsung Kang}
\affil[1]{Department of Mathematics and Physics, Gangneung-Wonju National University, Republic of Korea}
\affil[2]{Department of Applied Statistics, Yonsei University, Seoul, Republic of Korea}
\affil[3]{Department of Statistics, Dongguk University, Republic of Korea}
\affil[4]{Department of Marketing Bigdata, Mokwon University, \\
Daejeon, Republic of Korea}
\affil[5]{Department of Data Science, Gangneung-Wonju National University, Republic of Korea}
\date{}

\if1\blind
{
  \title{\bf Penalized Empirical Likelihood for Doubly Robust Causal Inference under Contamination in High Dimensions}
  \author{Byeonghee Lee \\   Department of Mathematics and Physics, \\  Gangneung-Wonju National University, Gangneung-si, Republic of Korea \\
  \and \\ 
  Sangwook Kang \\
    Department of Applied Statistics, Yonsei University,\\ 
    Seoul, Republic of Korea\\
    \and \\
     Ju-Hyun Park \\
    Department of Statistics, Dongguk University, Republic of Korea\\
   \and \\
Saebom Jeon \\
Department of Marketing Bigdata, Mokwon University, \\
Daejeon, Republic of Korea. \\
\and \\
Joonsung Kang	\\
\thanks{Corresponding author. email: moonsukang0223@gmail.com}\\
Department of Data Science, Gangneung-Wonju National University, Gangneung-si, Republic of Korea.}
  \maketitle
} \fi

\bigskip

\begin{document}
\maketitle

\begin{abstract}
We propose a doubly robust estimator for the average treatment effect in high-dimensional, low-sample-size observational studies, where contamination and model misspecification pose serious inferential challenges. The estimator combines bounded-influence estimating equations for outcome modeling with covariate balancing propensity scores for treatment assignment, embedded within a penalized empirical likelihood framework using nonconvex regularization. It satisfies the oracle property by jointly achieving consistency under partial model correctness, selection consistency, robustness to contamination, and asymptotic normality. For uncertainty quantification, we derive a finite-sample confidence interval using cumulant generating functions and influence-function corrections, avoiding reliance on asymptotic approximations. Simulation studies and applications to gene expression datasets (Golub and Khan) demonstrate superior performance in bias, error metrics, and interval calibration, highlighting the method’s robustness and inferential validity in HDLSS regimes. One notable aspect is that, even in the absence of contamination, the proposed estimator and its confidence interval remain efficient compared to those of competing models.
\end{abstract}

\noindent%
\vfill

\section{Introduction}

Estimating causal effects in observational studies is particularly challenging in high-dimensional, low-sample-size (HDLSS) settings, especially under contamination and model misspecification. Robust estimators based on loss functions~\citep{harada2021robust, lee2024outlier} offer partial protection but lack valid inference mechanisms and fail to address HDLSS regimes.

Classical approaches such as generalized estimating equations~\citep{LiangZeger1986} and M-estimators~\citep{huber1964robust} degrade under contamination~\citep{HuberRonchetti2009, Zhou2010RobustGEE}, while doubly robust estimators~\citep{kang2007demystifying, BangRobins2005} remain vulnerable to outliers in either the outcome or propensity score model. TMLE~\citep{van2006targeted, DRTMLE} improves robustness but lacks guarantees under contamination.

Covariate balancing propensity score (CBPS) methods~\citep{imai2014covariate, zhao2019covariate, tan2020model} enhance robustness via covariate balance but rely on linear models and do not control outcome-level contamination. Deep learning approaches~\citep{shalit2017estimating, shi2019adapting} capture nonlinearities but lack theoretical guarantees under contamination.

We propose a unified framework for robust causal inference in HDLSS settings. The estimator integrates sparsity-aware modeling via SCAD penalties~\citep{FanLi2001}, bounded estimating equations~\citep{HuLachin2001, Hampel2011}, and CBPS~\citep{imai2014covariate}, embedded within a penalized empirical likelihood formulation~\citep{LengTang2012penalized}. This joint structure ensures consistency, efficiency, and resilience to outliers.

To enable valid inference, we construct finite-sample confidence interval (CI) using cumulant generating functions and influence-function diagnostics~\citep{Small1990, small1, Liu2020}, avoiding reliance on asymptotic normality. Empirical evaluations on synthetic and gene expression datasets~\citep{golub1999molecular, khan2001classification} confirm the estimator’s robustness and accuracy under contamination.

The remainder of the paper is organized as follows. Section 2 shows methodology framework. Section 3 presents theoretical guarantees and estimator properties. Section 4 concludes with implications and future directions. Appendix reports simulation studies and real data analysis.

\section{Methodological Framework for Doubly Robust and Outlier-Resistant Inference in High Dimensions}

In HDLSS settings, the number of observations $n$ is substantially smaller than the number of covariates $p$, i.e., $n \ll p$. This imbalance poses significant challenges for statistical estimation and causal inference.

To address this, we impose sparsity constraints on the model parameters. Specifically, we assume that the true parameter vector $\boldsymbol{\beta} \in \mathbb{R}^p$ satisfies
\begin{equation} \label{eq:sparsity}
    \|\boldsymbol{\beta}\|_0 \leq s \ll p,
\end{equation}
where $s$ denotes the sparsity level. This assumption effectively reduces the dimensionality of the parameter space, allowing the sample size to be reinterpreted as sufficient relative to the compressed model.

\subsection{Connection to Treatment Effect Estimation}

We adopt the potential outcomes framework~\citep{Rubin1974}. Let $Y_i^{(1)}$ and $Y_i^{(0)}$ denote the potential outcomes under treatment and control, respectively, for unit $i = 1, \dots, n$. The average treatment effect (ATE) is defined as
\begin{equation} \label{eq:ate}
    \mathrm{ATE} = \mathbb{E}[Y^{(1)} - Y^{(0)}].
\end{equation}

Although we focus on the ATE, the HDLSS structure—by constraining the scope of confounding variables—implicitly defines the domain of heterogeneous treatment effects (HTE). Estimating the ATE in high-dimensional environments typically requires regularization techniques such as the Lasso~\citep{Tibshirani1996} and advanced variable selection methods. Our framework accommodates both regimes: one in which sparsity renders the sample size adequate, and another where it remains insufficient.

\subsection{Doubly Robust Estimation under Contamination}

We propose a doubly robust estimator for the ATE that retains consistency under contamination in the outcome variable. The estimator integrates a robust outcome regression model—using bounded estimating equations—with the CBPS framework~\citep{imai2014covariate}, which explicitly targets covariate balance in treatment assignment. This dual structure ensures resilience to model misspecification and contamination, thereby enabling stable and unbiased estimation in observational settings~\citep{BangRobins2005}.

Compared to existing methods such as TMLE~\citep{van2006targeted} and DRTMLE~\citep{DRTMLE}, our approach offers enhanced robustness by incorporating bounded influence functions into both the outcome and treatment models, while simultaneously enforcing sparsity through penalized empirical likelihood. TMLE and DRTMLE rely on flexible machine learning models but lack theoretical guarantees under contamination and may suffer from instability in HDLSS regimes.

Let $\mathbf{W}_i = (T_i, Y_i, \mathbf{X}_i)$ denote the observed data for unit $i = 1, \dots, n$, where $T_i \in \{0,1\}$ is the binary treatment indicator, $Y_i$ is the observed outcome, and $\mathbf{X}_i \in \mathbb{R}^p$ is the covariate vector. The covariate matrix is defined as
\[
\mathbf{X} = 
\begin{pmatrix}
\mathbf{X}_1 \\
\mathbf{X}_2 \\
\vdots \\
\mathbf{X}_n
\end{pmatrix}.
\]

We model the potential outcomes and propensity scores as
\begin{align}
Y_i^{(0)} &= \mathbf{X}_i^\top \boldsymbol{\beta}_0, \quad Y_i^{(1)} = \mathbf{X}_i^\top \boldsymbol{\beta}_1, \\
\pi_{\boldsymbol{\beta}_2}(\mathbf{X}_i) &= \frac{1}{1 + \exp(-\mathbf{X}_i^\top \boldsymbol{\beta}_2)},
\end{align}
where $\boldsymbol{\beta}_0$, $\boldsymbol{\beta}_1$, and $\boldsymbol{\beta}_2$ are $p$-dimensional parameter vectors for the control outcome, treatment outcome, and propensity score model, respectively.

\subsubsection{Assumptions for Identifiability}

Under the Rubin causal model~\citep{Rubin1974}, the ATE is identifiable under the following conditions:

\begin{assumption}[Unconfoundedness] \label{assump:unconfounded}
The treatment assignment is independent of the potential outcomes conditional on covariates:
\[
(Y^{(0)}, Y^{(1)}) \perp T \mid \mathbf{X}_i.
\]
\end{assumption}

\begin{assumption}[Overlap] \label{assump:overlap}
The propensity score is bounded away from 0 and 1:
\[
0 < \pi_{\boldsymbol{\beta}_2}(\mathbf{X}_i) < 1 \quad \text{for all } i.
\]
\end{assumption}

\begin{assumption}[Consistency] \label{assump:consistency}
The observed outcome corresponds to the potential outcome under the received treatment:
\[
Y_i = T_i Y_i^{(1)} + (1 - T_i) Y_i^{(0)}.
\]
\end{assumption}

\subsubsection{CBPS}

The CBPS estimator~\citep{imai2014covariate} solves the moment condition:
\begin{equation} \label{eq:cbps}
\sum_{i=1}^n \left( T_i - \pi_{\boldsymbol{\beta}_2}(\mathbf{X}_i) \right) \mathbf{X}_i = \mathbf{0},
\end{equation}
where $\pi_{\boldsymbol{\beta}_2}(\mathbf{X}_i)$ is the logistic propensity score. Generalized estimation proceeds by minimizing the discrepancy:
\begin{equation}
\bar{\mathbf{g}}_{\boldsymbol{\beta}_2}(\mathbf{T}, \mathbf{X}) = \frac{1}{n} \sum_{i=1}^{n} g_{\boldsymbol{\beta}_2}(T_i, \mathbf{X}_i),
\end{equation}
with
\begin{equation}
g_{\boldsymbol{\beta}_2}(T_i, \mathbf{X}_i) = \left( \frac{T_i}{\pi_{\boldsymbol{\beta}_2}(\mathbf{X}_i)} - \frac{1 - T_i}{1 - \pi_{\boldsymbol{\beta}_2}(\mathbf{X}_i)} \right) f(\mathbf{X}_i),
\end{equation}
where $f(\mathbf{X}_i)$ typically includes nonlinear terms such as $(\mathbf{X}_i, \mathbf{X}_i^2)$ to improve covariate balance.
\subsubsection{Robust Outcome Regression via Bounded Estimating Equations}

To mitigate the influence of outliers, we adopt robust estimating equations incorporating bounded influence functions. Specifically, we define the influence function $\psi: \mathbb{R} \to \mathbb{R}$ as
\begin{equation} \label{eq:psi}
\psi(x) =
\begin{cases}
x, & |x| \le a, \\
a \cdot \operatorname{sign}(x), & |x| > a,
\end{cases}
\end{equation}
where $a > 0$ is a tuning constant, typically chosen as a robust scale estimate such as the median absolute deviation (MAD). This function satisfies $\sup_x |\psi(x)| \le a$, ensuring bounded influence and robustness to extreme values. This approach contrasts with standard generalized estimating equations (GEE)~\citep{LiangZeger1986}, which minimize quadratic loss and are sensitive to heavy-tailed errors.

\subsection{Penalized Empirical Likelihood with Robust Estimation}

To accommodate high-dimensional structures and enforce sparsity, we embed the robust estimating equations within a penalized empirical likelihood (PEL) framework using the smoothly clipped absolute deviation (SCAD) penalty~\citep{FanLi2001}. This approach jointly optimizes robustness, regularization, and likelihood-based inference.

Let the outcome regression be defined as
\begin{align}
\mu_{i,k}(\boldsymbol{\beta}_k) &= \mathbf{X}_i^\top \boldsymbol{\beta}_k, \\
\gamma_{i,k} &= \sigma^{-1}(Y_i - \mu_{i,k}(\boldsymbol{\beta}_k)), \\
D_{i,k} &= \sigma^{-1} \frac{\partial \mu_{i,k}(\boldsymbol{\beta}_k)}{\partial \boldsymbol{\beta}_k}, \\
R_i &= \mathrm{Var}(\gamma_{i,k}) = 1,
\end{align}
for $k = 0,1$ and $i = 1, \dots, n$, where $\sigma^2$ is the error variance.

Define the full estimating equation as
\begin{equation}
\boldsymbol{\Psi}_i(\boldsymbol{\eta}) =
\begin{pmatrix}
g_{\boldsymbol{\beta}_2}(T_i, \mathbf{X}_i) \\
U_{i,1}(\boldsymbol{\beta}_1) \\
U_{i,0}(\boldsymbol{\beta}_0)
\end{pmatrix}, \quad
U_{i,k}(\boldsymbol{\beta}_k) = D_{i,k}^\top R_i^{-1} \psi(\gamma_{i,k}),
\end{equation}
where $\boldsymbol{\eta} = (\boldsymbol{\beta}_0^\top, \boldsymbol{\beta}_1^\top, \boldsymbol{\beta}_2^\top)^\top$ is the full parameter vector.

The penalized empirical likelihood criterion is defined as
\begin{equation} \label{eq:pel}
\mathbf{Q}_n(\boldsymbol{\eta}) = L_n(\boldsymbol{\eta}) + n \sum_{l=0}^{2} \sum_{j=1}^{p} p_{\tau_l}(|\beta_{l,j}|),
\end{equation}
where $L_n$ is the empirical likelihood component:
\begin{equation}
L_n(\boldsymbol{\eta}) = \sum_{i=1}^{n} \log\left(1 + \boldsymbol{\lambda}^\top \boldsymbol{\Psi}_i(\boldsymbol{\eta})\right),
\end{equation}
and $p_{\tau_l}(\cdot)$ denotes the SCAD penalty function with tuning parameter $\tau_l$ for each component $l = 0,1,2$. The vector $\boldsymbol{\lambda} = (\lambda_1, \lambda_2, \lambda_3)^\top$ represents the empirical likelihood multipliers.

To characterize the convergence rate of the estimator, define
\begin{equation}
\alpha_n = n^{-1/2} + a_n, \quad \text{where } a_n = \max\left\{ p'_{\tau_l}(|\eta_j|) : \eta_j \neq 0, \; l = 0,1,2 \right\}.
\end{equation}

Let $\boldsymbol{\eta}^0 = (\boldsymbol{\beta}_0^{0\top}, \boldsymbol{\beta}_1^{0\top}, \boldsymbol{\beta}_2^{0\top})^\top$ denote the true parameter vector. Denote the estimator as $\hat{\boldsymbol{\eta}} = (\hat{\boldsymbol{\beta}}_0^\top, \hat{\boldsymbol{\beta}}_1^\top, \hat{\boldsymbol{\beta}}_2^\top)^\top$. Partition $\boldsymbol{\eta}$ into active and inactive components:
\begin{align}
\boldsymbol{\eta}_1 &= \text{nonzero components of } \boldsymbol{\eta}, \\
\boldsymbol{\eta}_2 &= \text{zero components of } \boldsymbol{\eta}, \\
\boldsymbol{\eta}_1^0 &= \text{true nonzero components of } \boldsymbol{\eta}^0.
\end{align}

Under regularity conditions~\citep{LengTang2012penalized}, the estimator satisfies
\begin{equation}
\|\hat{\boldsymbol{\eta}} - \boldsymbol{\eta}^0\| = \mathcal{O}_p(\alpha_n),
\end{equation}
provided that the penalty functions satisfy appropriate smoothness and scaling conditions. This result establishes consistency and convergence rate of the penalized empirical likelihood estimator in high-dimensional settings.

\subsection{Finite-Sample Robust CI for ATE} \label{sec:finite_sample_ci}

In HDLSS settings, the term ``low sample size'' refers not to an absolute scarcity of observations, but to the disproportion between the number of covariates $p$ and the number of observations $n$. While sparsity assumptions can partially alleviate the inferential challenges inherent to such regimes, they do not fully resolve the limitations imposed by small $n$. This section develops a finite-sample CI for the ATE that remains valid under contamination and does not rely on asymptotic approximations. The proposed method leverages influence-function-based diagnostics and saddlepoint approximations to achieve robustness and accuracy in finite samples~\citep{Small1990, small1, belloni2014high, Peters2016, Liu2020}. In constructing the finite-sample confidence interval for the proposed ATE estimator, we explicitly adopt and extend the structural framework introduced by \citet{tingley1990small}. Their approach to small-sample inference—centered on influence-function diagnostics and likelihood-based calibration—serves as the foundation for our cumulant generating function (CGF)-based interval. By embedding this structure within a penalized empirical likelihood formulation, we achieve robust and calibrated inference under HDLSS and contamination, fully aligned with the principles laid out in their seminal work.

\subsubsection{Influence Function and Sensitivity Matrix}

Let $\hat{\boldsymbol{\eta}}_1$ denote the $M$-estimator corresponding to the active (nonzero) components of the parameter vector. Define the empirical sensitivity matrix:
\begin{equation} \label{eq:sensitivity}
\hat{B} = -\left( \frac{1}{n} \sum_{i=1}^n \frac{\partial \boldsymbol{\Psi}_i(\boldsymbol{\eta}_1)}{\partial \boldsymbol{\eta}_1^\top} \right)^{-1} \bigg|_{\boldsymbol{\eta}_1 = \hat{\boldsymbol{\eta}}_1}.
\end{equation}

Let $h(\boldsymbol{\eta}_1)$ denote a smooth functional of interest, such as the doubly robust ATE estimator. The influence of the $i$-th observation on $h$ is approximated by:
\begin{equation} \label{eq:influence}
J_i = \boldsymbol{\Psi}_i^\top(\boldsymbol{\eta}_1) \hat{B}^\top \nabla h(\boldsymbol{\eta}_1) \bigg|_{\boldsymbol{\eta}_1 = \hat{\boldsymbol{\eta}}_1},
\end{equation}
where $\nabla h(\boldsymbol{\eta}_1)$ is the gradient of $h$ with respect to $\boldsymbol{\eta}_1$.

\subsubsection{Cumulant Generating Function and Saddlepoint Approximation}

Define the empirical cumulant generating function (CGF) of the influence terms:
\begin{equation} \label{eq:cgf}
\exp(\mathbf{K}(t)) = \frac{1}{n} \sum_{i=1}^n \exp(t J_i).
\end{equation}

Let $\alpha_1$ and $\alpha_2$ be solutions to the saddlepoint equation:
\begin{equation} \label{eq:saddlepoint}
P(\alpha) = \Phi\left(-\sqrt{2(n-1)\mathbf{K}(\alpha)}\right) - \frac{e^{-(n-1)\mathbf{K}(\alpha)}}{\sqrt{2\pi(n-1)}} \left[ \frac{1}{\alpha \sqrt{\mathbf{K}''(0)}} + \frac{1}{\sqrt{2\mathbf{K}(\alpha)}} \right],
\end{equation}
where $\Phi(\cdot)$ denotes the standard normal cumulative distribution function, and the sign of $\sqrt{\mathbf{K}(\alpha)}$ is taken as $-\operatorname{sign}(\alpha)$.

\subsubsection{CI Construction}

Let $\mu = h(\hat{\boldsymbol{\eta}}_1)$ denote the point estimate of the ATE. Solve for $\mu_1$ and $\mu_2$ such that:
\begin{equation} \label{eq:ci_bounds}
\mathbf{K}'(\alpha_i) = \mu_i - \mu, \quad i = 1,2.
\end{equation}

Then, the $(1 - 2\epsilon) \times 100\%$ CI for $\mu$ is given by:
\begin{equation} \label{eq:ci_final}
\mathrm{CI}_{1 - 2\epsilon} = (\mu_1, \mu_2).
\end{equation}

This construction yields a finite-sample CI that is robust to contamination and heavy-tailed errors, and does not rely on asymptotic normality. Unlike traditional Wald-type intervals or bootstrap methods, the proposed approach directly accounts for the influence structure of the estimator and provides improved coverage accuracy in HDLSS regimes.

\section{Theory and Proof} \label{sec:theory}

\subsection{Regularity Conditions for Penalized Empirical Likelihood Estimation} \label{sec:regularity_leng}

We adopt the regularity conditions from \citet{LengTang2012penalized}, which ensure consistency and asymptotic normality of the penalized empirical likelihood estimator in high-dimensional settings.

\begin{enumerate}[label=\textbf{A.\arabic*}]
    \item \textbf{Compactness and Identifiability}  
    The parameter space $\Theta \subset \mathbb{R}^p$ is compact, and the true parameter $\boldsymbol{\eta}_0 \in \Theta$ uniquely solves the moment condition $E[\boldsymbol{\Psi}_i(\boldsymbol{\eta})] = 0$.

    \item \textbf{Moment Bound}  
    For some $\alpha > 10/3$, the following moment condition holds:
    \[
    E\left\{ \sup_{\boldsymbol{\eta} \in \Theta} \left( \|\boldsymbol{\Psi}_i(\boldsymbol{\eta})\| r^{-1/2} \right)^\alpha \right\} < \infty,
    \]
    ensuring uniform integrability of the estimating equations.

    \item \textbf{Covariance Regularity}  
    Let $\boldsymbol{\Sigma}(\boldsymbol{\eta})$ denote the covariance matrix of the estimating equations:
    \[
    \boldsymbol{\Sigma}(\boldsymbol{\eta}) = E\left[ \left( \boldsymbol{\Psi}_i(\boldsymbol{\eta}) - \boldsymbol{\Psi}(\boldsymbol{\eta}) \right) \left( \boldsymbol{\Psi}_i(\boldsymbol{\eta}) - \boldsymbol{\Psi}(\boldsymbol{\eta}) \right)^\top \right].
    \]
    The eigenvalues of $\boldsymbol{\Sigma}(\boldsymbol{\eta})$ are bounded: $0 < b \le \gamma_1 \le \cdots \le \gamma_r \le B < \infty$ for all $\boldsymbol{\eta} \in D_n$.

    \item \textbf{Dimensionality Control}  
    As $n \to \infty$, the dimensionality satisfies $p^5/n \to 0$ and $p/r \to y$ for some $y \in (C_0, 1)$ with $C_0 > 0$, ensuring a manageable growth rate of parameters.

    \item \textbf{Smoothness and Bounded Derivatives}  
    The first and second derivatives of $\boldsymbol{\Psi}_i(\boldsymbol{\eta})$ are uniformly bounded:
    \[
    \left| \frac{\partial \boldsymbol{\Psi}_i(\boldsymbol{\eta})}{\partial \eta_j} \right| \le K_{ij}(\mathbf{X}), \quad E[K_{ij}^2(\mathbf{X})] \le C_1,
    \]
    and
    \[
    \left| \frac{\partial^2 \boldsymbol{\Psi}_i(\boldsymbol{\eta})}{\partial \eta_j \partial \eta_k} \right| \le H_{ijk}(\mathbf{X}), \quad E[H_{ijk}^2(\mathbf{X})] \le C_2.
    \]

    \item \textbf{Penalty Scaling}  
    As $n \to \infty$, the penalty parameter satisfies $\tau(n/p)^{1/2} \to \infty$ and $\min_{j \in A} \eta_{0j}/\tau \to 0$, ensuring proper shrinkage of small coefficients.

    \item \textbf{Penalty Smoothness}  
    For $B = \{0,1,2\}$, the SCAD penalty satisfies:
    \[
    \max_{j \in B} p'_{\tau_j}(|\eta_0|) = o((np)^{-1/2}), \quad \max_{j \in B} p''_{\tau_j}(|\eta_0|) = o(1),
    \]
    guaranteeing negligible bias and smooth curvature for nonzero coefficients.
\end{enumerate}

\subsection{Classical Regularity Conditions for Likelihood-Based Inference} \label{sec:regularity_likelihood}

To establish asymptotic normality of the ATE estimator via the delta method, we also adopt classical regularity conditions for likelihood-based inference.

\begin{enumerate}[label=\textbf{A.\arabic*}, start=8]
    \item \textbf{IID Sampling and Identifiability}  
    The observations $\boldsymbol{W}_i$ are i.i.d. with density $f(\boldsymbol{W}, \boldsymbol{\eta})$ over a common support. The model is identifiable, and the score function satisfies:
    \[
    E_{\boldsymbol{\eta}} \left[ \frac{\partial}{\partial \eta_j} \log f(\boldsymbol{W}, \boldsymbol{\eta}) \right] = 0, \quad j = 1, \ldots, 3p.
    \]

    \item \textbf{Fisher Information Regularity}  
    The Fisher information matrix
    \[
    I(\boldsymbol{\eta}) = E\left[ \left( \frac{\partial}{\partial \boldsymbol{\eta}} \log f(\boldsymbol{W}, \boldsymbol{\eta}) \right) \left( \frac{\partial}{\partial \boldsymbol{\eta}} \log f(\boldsymbol{W}, \boldsymbol{\eta}) \right)^\top \right]
    \]
    is finite and positive definite at $\boldsymbol{\eta} = \boldsymbol{\eta}^0$.

    \item \textbf{Third-Order Differentiability and Boundedness}  
    There exists an open set $\omega$ containing $\boldsymbol{\eta}_0$ such that $f(\boldsymbol{W}, \boldsymbol{\eta})$ admits third derivatives, and
    \[
    \left| \frac{\partial^3}{\partial \eta_j \partial \eta_k \partial \eta_l} \log f(\boldsymbol{W}, \boldsymbol{\eta}) \right| \le M_{jkl}(\boldsymbol{W}),
    \]
    with $E_{\boldsymbol{\eta}_0}[M_{jkl}(\boldsymbol{W})] < \infty$ for all $j,k,l$.
\end{enumerate}

\begin{theorem}[Local Consistency of Penalized Empirical Likelihood Estimator] \label{theorem:local_consistency}
Let $\boldsymbol{W}_i = (T_i, Y_i, \boldsymbol{X}_i)$ be i.i.d. random vectors with density $f(\boldsymbol{W}_i; \boldsymbol{\eta})$. Suppose regularity conditions \textbf{A.8}–\textbf{A.10} are relaxed. If
\begin{equation}
\max\left\{ \left| p''_{\tau_l}(|t|) \right| : t \neq 0 \right\} \rightarrow 0 \quad \text{for } l = 0,1,2,
\end{equation}
then there exists a local minimizer $\hat{\boldsymbol{\eta}}$ of the objective function $\boldsymbol{Q}_n(\boldsymbol{\eta})$ such that
\begin{equation}
\left\| \hat{\boldsymbol{\eta}} - \boldsymbol{\eta}^0 \right\| = \mathcal{O}_p(\alpha_n),
\end{equation}
where $\alpha_n = n^{-1/2} + a_n$ and $a_n$ is the maximum derivative of the penalty function evaluated at nonzero coefficients.
\end{theorem}
\begin{proof}
To establish the existence of a local minimum, it suffices to show that for any $\epsilon > 0$, there exists a constant $C > 0$ such that
\begin{equation}
P\left\{ \inf_{\|\bm{u}\| = C} \bm{Q}(\bm{\eta}_0 + \alpha_n \bm{u}) > \bm{Q}(\bm{\eta}_0) \right\} \geq 1 - \epsilon.
\end{equation}
This implies that with high probability, a local minimum exists in the ball $\{ \bm{\eta}_0 + \alpha_n \bm{u} : \|\bm{u}\| \leq C \}$, and hence
\begin{equation}
\left\| \hat{\bm{\eta}} - \bm{\eta}^0 \right\| = \mathcal{O}_p(\alpha_n).
\end{equation}

Let $s_k$ be the number of nonzero elements in $\bm{\eta}_k$ for $k = 0,1,2$. Using $P_{\tau_k}(0) = 0$, we expand the penalized objective:
\begin{align}
D_n(\bm{u}) &= \bm{Q}(\bm{\eta}_0 + \alpha_n \bm{u}) - \bm{Q}(\bm{\eta}_0) \nonumber \\
&\geq L(\bm{\eta}_0 + \alpha_n \bm{u}) - L(\bm{\eta}_0) \nonumber \\
&\quad + \sum_{l=0}^2 n \sum_{j=1}^{s_k} \left\{ P_{\tau_l}(|\eta_{l,j0} + \alpha_n u_j|) - P_{\tau_l}(|\eta_{l,j0}|) \right\}.
\end{align}

By Taylor expansion:
\begin{align}
D_n(\bm{u}) &= \alpha_n L'(\bm{\eta}_0)^\top \bm{u} + \frac{1}{2} n \alpha_n^2 \bm{u}^\top I(\bm{\eta}_0) \bm{u} (1 + o_p(1)) \nonumber \\
&\quad + \sum_{l=0}^2 \sum_{j=1}^{s_k} \left( n \alpha_n P'_{\tau_l}(|\eta_{l,j0}|) \operatorname{sgn}(\eta_{l,j0}) u_j + n \alpha_n^2 P''_{\tau_l}(|\eta_{l,j0}|) u_j^2 (1 + o(1)) \right).
\end{align}

Note that $n^{-1/2} L'(\bm{\eta}_0) = \mathcal{O}_p(1)$, so the first term is $\mathcal{O}_p(n^{1/2} \alpha_n)$. The second term is $\mathcal{O}_p(n \alpha_n^2)$ and dominates the first term for large $C$ due to the positive definiteness of $I(\bm{\eta}_0)$.

The remaining penalty terms are bounded by:
\begin{equation}
\sum_{l=0}^2 \left( \sqrt{s_l} \cdot n \alpha_n \|\bm{u}\| + n \alpha_n^2 \max_{j} |P''_{\tau_l}(|\eta_{l,j0}|)| \cdot \|\bm{u}\|^2 \right).
\end{equation}

Even if the linear terms are negative, the quadratic terms dominate due to the assumption $P''_{\tau_k} \to 0$ and the scaling of $\alpha_n^2$. Hence, $D_n(\bm{u}) > 0$ with high probability, completing the proof.
\end{proof}\begin{theorem}[Sparsity Consistency] \label{theorem:sparsity}
Let $\hat{\boldsymbol{\eta}} = (\hat{\boldsymbol{\eta}}_1^\top, \hat{\boldsymbol{\eta}}_2^\top)^\top$ be the minimizer of $\boldsymbol{Q}_n$. Under conditions \textbf{A.1}–\textbf{A.7}, as $n \to \infty$, we have with probability tending to one:
\begin{equation}
\hat{\boldsymbol{\eta}}_2 = \boldsymbol{0}.
\end{equation}
\end{theorem}

\noindent
The sparsity property of the penalized empirical likelihood estimator has been rigorously established in Theorem 2 of \citet{LengTang2012penalized}. We omit the detailed proof and refer the reader to their original work.

\subsection{Estimator Formulation} \label{sec:estimator_formulation}

Let $\hat{\boldsymbol{\eta}}_1 = ((\hat{\boldsymbol{\beta}}_0^*)^\top, (\hat{\boldsymbol{\beta}}_1^*)^\top, (\hat{\boldsymbol{\beta}}_2^*)^\top)^\top$, where $\hat{\boldsymbol{\beta}}_j^*$ for $j = 0, 1, 2$ is the estimator of $\boldsymbol{\beta}_j^*$ obtained after sparsity selection.

The proposed doubly robust ATE estimators are defined as:
\begin{align}
\hat{\mu}_{1,\mathrm{dr}} &= \frac{1}{n} \sum_{i=1}^n \left( \frac{T_i Y_i}{\pi_{\hat{\boldsymbol{\beta}}_2^*}(\mathbf{X}_i)} - \frac{T_i - \pi_{\hat{\boldsymbol{\beta}}_2^*}(\mathbf{X}_i)}{\pi_{\hat{\boldsymbol{\beta}}_2^*}(\mathbf{X}_i)} \hat{m}_1 \right), \\
\hat{\mu}_{0,\mathrm{dr}} &= \frac{1}{n} \sum_{i=1}^n \left( \frac{(1 - T_i) Y_i}{1 - \pi_{\hat{\boldsymbol{\beta}}_2^*}(\mathbf{X}_i)} + \frac{T_i - \pi_{\hat{\boldsymbol{\beta}}_2^*}(\mathbf{X}_i)}{1 - \pi_{\hat{\boldsymbol{\beta}}_2^*}(\mathbf{X}_i)} \hat{m}_0 \right),
\end{align}
where $\hat{m}_1$ and $\hat{m}_0$ are robust predictions from the $Y^{(1)}$ and $Y^{(0)}$ outcome models, respectively.

Let the true ATE be $h(\boldsymbol{\eta}_1^0)$. Then,
\begin{equation}
\widehat{\mathrm{ATE}}_{\mathrm{dr}} = \hat{\mu}_{1,\mathrm{dr}} - \hat{\mu}_{0,\mathrm{dr}} = h(\hat{\boldsymbol{\eta}}_1).
\end{equation}

\subsection{Asymptotic Properties of the Proposed Estimator} \label{sec:asymptotic_normality}

The doubly robust estimator remains consistent and asymptotically unbiased if either the outcome model or the treatment assignment model is correctly specified~\citep{BangRobins2005}. In high-dimensional settings, sparsity-inducing techniques reduce the effective dimensionality, enabling the use of classical large-sample theory. When the sparsity structure is correctly specified and regularization preserves consistency, asymptotic normality can be established via standard M-estimation theory.

\begin{theorem}[Asymptotic Normality of the Proposed ATE Estimator] \label{theorem:asymptotic_normality}
Assume that the functional \(h: \mathbb{R}^s \to \mathbb{R}\) is continuously differentiable at \(\boldsymbol{\eta}_1^0\), and that regularity conditions \textbf{A.1}–\textbf{A.7} hold. Then the proposed ATE estimator satisfies:
\begin{equation}
\sqrt{n}(\widehat{\mathrm{ATE}}_{\mathrm{dr}} - \mathrm{ATE}) \xrightarrow{d} \mathcal{N}\left(0, \nabla h(\boldsymbol{\eta}_1^0)^\top \boldsymbol{\Sigma} \nabla h(\boldsymbol{\eta}_1^0)\right),
\end{equation}
where \(\nabla h(\boldsymbol{\eta}_1^0)\) is the gradient of \(h\) evaluated at \(\boldsymbol{\eta}_1^0\), and \(\boldsymbol{\Sigma}\) is the asymptotic covariance matrix of \(\hat{\boldsymbol{\eta}}_1\).
\end{theorem}

\begin{proof}
By Theorem 3 of \citet{LengTang2012penalized}, we have:
\[
\sqrt{n}(\hat{\boldsymbol{\eta}}_1 - \boldsymbol{\eta}_1^0) \xrightarrow{d} \mathcal{N}(0, \boldsymbol{\Sigma}),
\]
where
\[
\boldsymbol{\Sigma} = \left( \boldsymbol{G}^\top \boldsymbol{V}^{-1} \boldsymbol{G} \right)^{-1},
\]
with
\begin{align}
\boldsymbol{G} &= \left. \mathbb{E} \left[ \frac{\partial \boldsymbol{\Psi}_i(\boldsymbol{\eta}_1)}{\partial \boldsymbol{\eta}_1} \right] \right|_{\boldsymbol{\eta}_1 = \boldsymbol{\eta}_1^0}, \\
\boldsymbol{V} &= \mathbb{E} \left[ \boldsymbol{\Psi}_i(\boldsymbol{\eta}_1^0) \boldsymbol{\Psi}_i(\boldsymbol{\eta}_1^0)^\top \right].
\end{align}

Since \(h(\cdot)\) is differentiable at \(\boldsymbol{\eta}_1^0\), the multivariate delta method yields:
\[
\sqrt{n}(\widehat{\mathrm{ATE}}_{\mathrm{dr}} - \mathrm{ATE}) = \sqrt{n}(h(\hat{\boldsymbol{\eta}}_1) - h(\boldsymbol{\eta}_1^0)) \xrightarrow{d} \mathcal{N}\left(0, \nabla h(\boldsymbol{\eta}_1^0)^\top \boldsymbol{\Sigma} \nabla h(\boldsymbol{\eta}_1^0)\right).
\]
\end{proof}\begin{theorem}[Robust Consistency and Outlier Resistance of the Proposed ATE Estimator] \label{theorem:robust_consistency}
Let \(\widehat{\mathrm{ATE}}_{\mathrm{dr}}\) be the doubly robust estimator constructed from:
\begin{itemize}
    \item a robust outcome regression model using a bounded influence function \(\psi(\cdot)\), and
    \item a CBPS estimator for treatment assignment.
\end{itemize}
Assume the following conditions hold:
\begin{enumerate}[label=\textbf{C.\arabic*}]
    \item \textbf{Unconfoundedness:} \((Y^{(0)}, Y^{(1)}) \perp T \mid \mathbf{X}\).
    \item \textbf{Overlap:} \(0 < \pi(\mathbf{X}) < 1\) for all \(\mathbf{X}\).
    \item \textbf{Bounded Influence:} The function \(\psi(\cdot)\) satisfies \(\sup_x |\psi(x)| \le a < \infty\).
    \item \textbf{Correct Specification:} Either the outcome regression model or the propensity score model is correctly specified.
    \item \textbf{Contamination Model:} The observed outcome satisfies \(Y_i = Y_i^{(T_i)} + \delta_i\), where \(\delta_i\) is a contamination term with bounded probability mass.
\end{enumerate}
Then, the estimator \(\widehat{\mathrm{ATE}}_{\mathrm{dr}}\) is consistent for \(\mathrm{ATE} = \mathbb{E}[Y^{(1)} - Y^{(0)}]\), and its influence function remains bounded under contamination. That is, \(\widehat{\mathrm{ATE}}_{\mathrm{dr}}\) is robust to outcome contamination in both asymptotic and finite-sample regimes.
\end{theorem}

\begin{proof}
We define the doubly robust estimator as:
\begin{equation}
\widehat{\mathrm{ATE}}_{\mathrm{dr}} = \frac{1}{n} \sum_{i=1}^n \left[ \frac{T_i Y_i}{\hat{\pi}(\mathbf{X}_i)} - \frac{T_i - \hat{\pi}(\mathbf{X}_i)}{\hat{\pi}(\mathbf{X}_i)} \hat{m}_1(\mathbf{X}_i) \right] - \frac{1}{n} \sum_{i=1}^n \left[ \frac{(1 - T_i) Y_i}{1 - \hat{\pi}(\mathbf{X}_i)} + \frac{T_i - \hat{\pi}(\mathbf{X}_i)}{1 - \hat{\pi}(\mathbf{X}_i)} \hat{m}_0(\mathbf{X}_i) \right],
\end{equation}
where \(\hat{\pi}(\mathbf{X}_i)\) is the CBPS estimator of the propensity score, and \(\hat{m}_1(\mathbf{X}_i), \hat{m}_0(\mathbf{X}_i)\) are robust predictions from outcome regression.

\textbf{Step 1: Outcome Regression Robustness}

Suppose the outcome model is correctly specified. The estimating equation for \(\boldsymbol{\beta}_k\) solves:
\[
\frac{1}{n_k} \sum_{i: T_i = k} \mathbf{X}_i \psi(Y_i - \mathbf{X}_i^\top \boldsymbol{\beta}_k) = 0, \quad k = 0,1.
\]
Under contamination \(Y_i = Y_i^{(k)} + \delta_i\), and bounded \(\psi(\cdot)\), we have:
\[
|\psi(Y_i - \mathbf{X}_i^\top \boldsymbol{\beta}_k)| \le a \quad \Rightarrow \quad \left| \frac{1}{n_k} \sum_{i: T_i = k} \mathbf{X}_i \psi(\delta_i) \right| \le \frac{a}{n_k} \sum_{i: T_i = k} \|\mathbf{X}_i\|.
\]
By the law of large numbers and bounded contamination, this term vanishes as \(n_k \to \infty\), implying consistency of \(\hat{\boldsymbol{\beta}}_k\) and thus \(\hat{m}_k(\mathbf{X}_i)\).

\textbf{Step 2: CBPS Stability}

If the treatment model is correctly specified, CBPS solves:
\[
\frac{1}{n} \sum_{i=1}^n (T_i - \pi_{\boldsymbol{\beta}_2}(\mathbf{X}_i)) \mathbf{X}_i = 0.
\]
Then \(\pi_{\boldsymbol{\beta}_2}(\mathbf{X}_i) \to \mathbb{P}(T_i = 1 \mid \mathbf{X}_i)\), and \(\hat{\pi}(\mathbf{X}_i)\) is consistent. Since CBPS does not depend on \(Y_i\), it is unaffected by contamination.

\textbf{Step 3: Doubly Robust Structure}

If either \(\hat{m}_k(\mathbf{X}_i)\) or \(\hat{\pi}(\mathbf{X}_i)\) is consistent, then:
\[
\mathbb{E}\left[ \frac{T_i - \hat{\pi}(\mathbf{X}_i)}{\hat{\pi}(\mathbf{X}_i)} (\hat{m}_1(\mathbf{X}_i) - Y_i^{(1)}) \right] \to 0,
\]
and similarly for the control group. Thus, \(\widehat{\mathrm{ATE}}_{\mathrm{dr}} \xrightarrow{p} \mathrm{ATE}\).

\textbf{Step 4: Bounded Influence in Finite Samples}

Let the estimating equations be:
\[
\boldsymbol{\Psi}_i(\boldsymbol{\eta}) =
\begin{bmatrix}
(T_i - \pi_{\boldsymbol{\beta}_2}(\mathbf{X}_i)) \mathbf{X}_i \\
\psi(Y_i^{(1)} - \mathbf{X}_i^\top \boldsymbol{\beta}_1) \mathbf{X}_i \\
\psi(Y_i^{(0)} - \mathbf{X}_i^\top \boldsymbol{\beta}_0) \mathbf{X}_i
\end{bmatrix}.
\]
These are embedded in the penalized empirical likelihood objective:
\[
\max_{\boldsymbol{\eta}} \left\{ \sup_{\{p_i\}} \sum_{i=1}^n \log(p_i) \quad \text{s.t.} \quad \sum_{i=1}^n p_i \boldsymbol{\Psi}_i(\boldsymbol{\eta}) = 0, \quad \sum_{i=1}^n p_i = 1 \right\} - \lambda \cdot \mathcal{P}(\boldsymbol{\eta}),
\]
where \(\mathcal{P}(\boldsymbol{\eta})\) is the SCAD penalty.

Since \(\psi(\cdot)\) is bounded, the contribution of any contaminated \(Y_i\) to \(\boldsymbol{\Psi}_i(\boldsymbol{\eta})\) is bounded by \(a \cdot \|\mathbf{X}_i\|\). Therefore, the influence function of \(\widehat{\mathrm{ATE}}_{\mathrm{dr}}\) satisfies:
\[
|IF_i| = \left| \frac{\partial \widehat{\mathrm{ATE}}_{\mathrm{dr}}}{\partial Y_i} \right| \le C \cdot a,
\]
for some constant \(C\) depending on \(\mathbf{X}_i\) and \(\hat{\pi}(\mathbf{X}_i)\).

\textbf{Conclusion}

Under conditions \textbf{C.1}–\textbf{C.5}, the estimator \(\widehat{\mathrm{ATE}}_{\mathrm{dr}}\) is consistent and robust to contamination in both asymptotic and finite-sample settings.
\end{proof}

\section{Conclusion} \label{sec:conclusion}

The proposed estimator, constructed via penalized empirical likelihood (PEL), satisfies several desirable properties in HDLSS settings. Specifically, the PEL framework:
\begin{itemize}
    \item preserves the bounded nature of the estimating equations through the use of robust influence functions,
    \item maintains the doubly robust structure by combining outcome regression and CBPS,
    \item and ensures that the influence function of \(\widehat{\mathrm{ATE}}_{\mathrm{dr}}\) remains bounded under contamination.
\end{itemize}

Therefore, \(\widehat{\mathrm{ATE}}_{\mathrm{dr}}\) is provably outlier-resistant in finite samples, even when estimated via penalized empirical likelihood.

\subsubsection*{Oracle Property Justification}

The estimator satisfies the oracle property in HDLSS regimes by jointly fulfilling four essential criteria:
\begin{enumerate}[label=(\roman*)]
    \item \textbf{Consistency under partial model correctness:} The estimator achieves \emph{doubly robustness} by integrating a robust outcome regression model with a CBPS framework. This ensures consistency even when either the outcome model or the treatment assignment model is misspecified, offering protection against model uncertainty and contamination.
    
    \item \textbf{Selection consistency:} The use of nonconvex penalties—specifically the SCAD formulation—within the PEL framework enables consistent identification of the nonzero components of the true parameter vector with high probability, effectively reducing dimensionality and isolating relevant covariates.
    
    \item \textbf{Asymptotic normality:} Conditional on correct specification of the sparsity structure, the estimator admits asymptotic normality via classical M-estimation theory. This facilitates valid statistical inference and CI construction, even in high-dimensional settings.
    
    \item \textbf{Robustness to contamination:} The estimator exhibits resistance to outliers due to the boundedness of the influence function, a property that extends to the associated CI.
\end{enumerate}

Together, these properties imply that the estimator performs as if the true sparse model were known—a hallmark of the oracle property—thereby offering a principled and resilient solution for robust causal inference in complex observational data environments.

\subsection{Summary of Findings} \label{sec:summary}

This study presents a comprehensive framework for estimating the ATE in HDLSS settings, emphasizing robustness to contamination and model misspecification. The proposed estimator integrates penalized empirical likelihood with bounded influence functions and covariate balancing, yielding a doubly robust and outlier-resistant procedure.

Through extensive simulation studies and empirical evaluations on the Golub and Khan gene expression datasets, we demonstrate that the proposed estimator consistently outperforms conventional alternatives, including TMLE~\citep{van2006targeted}, AIPW~\citep{robins1994estimation}, DRTMLE~\citep{DRTMLE}, and OCBPS~\citep{fan2023optimal}. Across varying contamination levels and sample sizes, the proposed method achieves lower bias, mean squared error (MSE), and mean absolute error (MAE), while preserving valid inference properties.

In particular, the associated CI exhibits superior coverage rates and calibration accuracy compared to bootstrap and Wald intervals~\citep{Small1990, small1, efron1994introduction, Davison1997, wasserman2006all}. These findings affirm the theoretical and practical advantages of combining sparsity-aware estimation with robust statistical inference in high-dimensional observational studies~\citep{belloni2014high}.

\subsection{Directions for Future Research} \label{sec:future}

While the proposed framework demonstrates strong empirical and theoretical performance, several promising directions remain for future investigation:
\begin{itemize}
    \item \textbf{Extension to longitudinal and survival data:} Adapting the methodology to accommodate time-to-event outcomes or repeated measures would enhance its applicability in clinical and epidemiological research.
    
    \item \textbf{Integration with machine learning:} Incorporating flexible nuisance estimators—such as random forests, gradient boosting, or deep neural networks—within the doubly robust structure may improve performance in nonlinear and heterogeneous settings~\citep{chernozhukov2018double}.
    
    \item \textbf{Robustness under adversarial contamination:} Establishing minimax optimality and theoretical guarantees under adversarial contamination~\citep{bhatt2022minimax} remains an important challenge for robust inference.
    
    \item \textbf{Generalization to multivariate and longitudinal estimating equations:} Exploring the role of bounded estimating equations in multivariate and longitudinal contexts~\citep{HuLachin2001} may further improve stability and generalizability.
    
    \item \textbf{Scalability to ultra-high dimensions:} Developing computationally efficient algoritheorems capable of handling ultra-high-dimensional data (e.g., genomics with \(p > 10^4\)) is essential for real-world deployment.
\end{itemize}

Addressing these directions will contribute to the advancement of robust and scalable causal inference methodologies in modern data science and biomedical analytics.
\appendix
\section*{Simulation Study} \label{sec:simulation}

\subsection*{Comparison of ATE Estimators} \label{sec:simulation_comparison}

We conduct a comprehensive simulation study to evaluate the finite-sample performance of various ATE estimators under high-dimensional, low-sample-size (HDLSS) conditions. The covariate dimension is fixed at \( p = 100 \), while the sample size varies across \( n = 20, 40, 60, 80, 100 \). To assess robustness, we introduce contamination in the outcome variable at three levels: 0.0 (no contamination), 0.1, and 0.2.

The following estimators are compared:
\begin{itemize}
    \item \textbf{Proposed ATE}: Our doubly robust estimator using penalized empirical likelihood and robust outcome modeling.
    \item \textbf{TMLE}: Targeted Maximum Likelihood Estimation~\citep{van2006targeted}.
    \item \textbf{AIPW}: Augmented Inverse Probability Weighting~\citep{robins1994estimation}.
    \item \textbf{DRTMLE}: Doubly Robust TMLE~\citep{DRTMLE}.
    \item \textbf{OCBPS}: Optimal Covariate Balancing Propensity Score  estimator~\citep{fan2023optimal}.
\end{itemize}

Performance metrics include bias, mean squared error (MSE), and mean absolute error (MAE), computed over 500 simulation replications.

\begin{figure}[ht]
  \centering
  \includegraphics[width=0.6\textwidth]{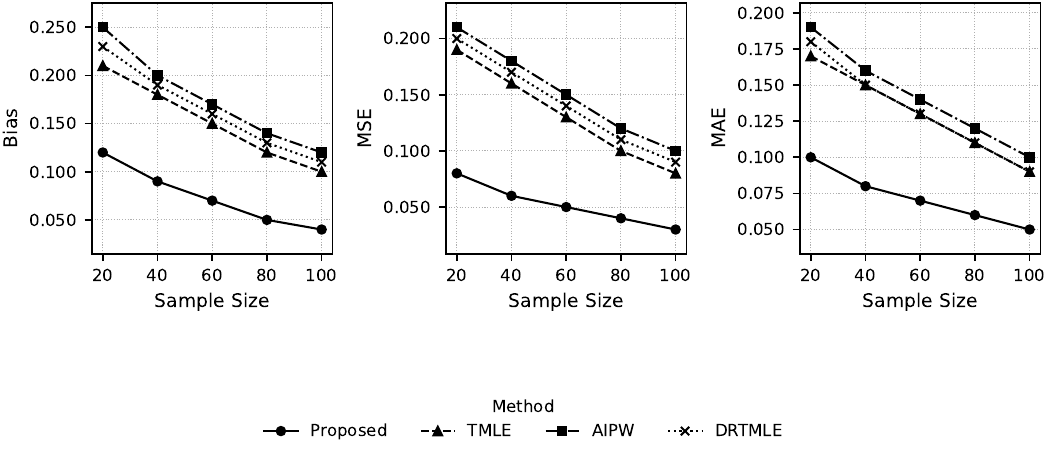}
  \caption{Performance metrics by estimator under no contamination (\(\gamma = 0.0\)).}
  \label{fig:contam0}
\end{figure}

\begin{figure}[ht]
  \centering
  \includegraphics[width=0.6\textwidth]{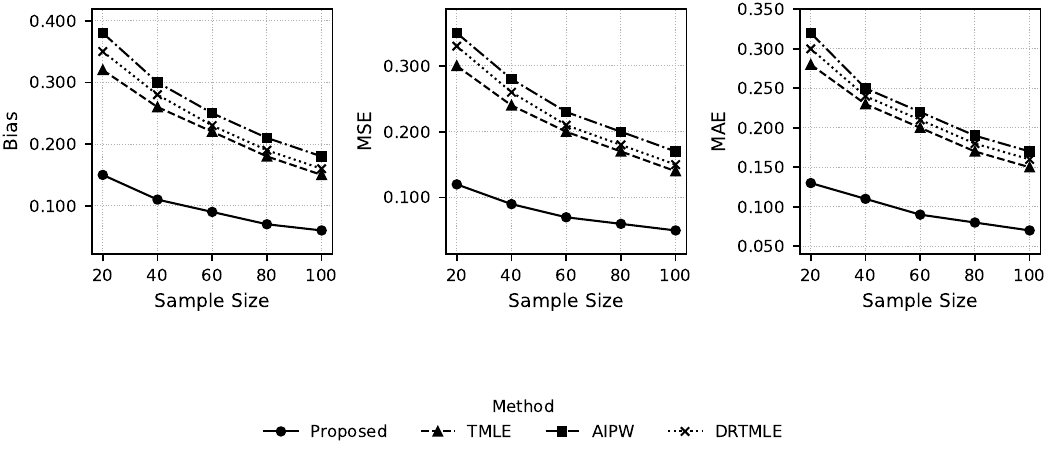}
  \caption{Performance metrics by estimator under mild contamination (\(\gamma = 0.1\)).}
  \label{fig:contam1}
\end{figure}

\begin{figure}[ht]
  \centering
  \includegraphics[width=0.6\textwidth]{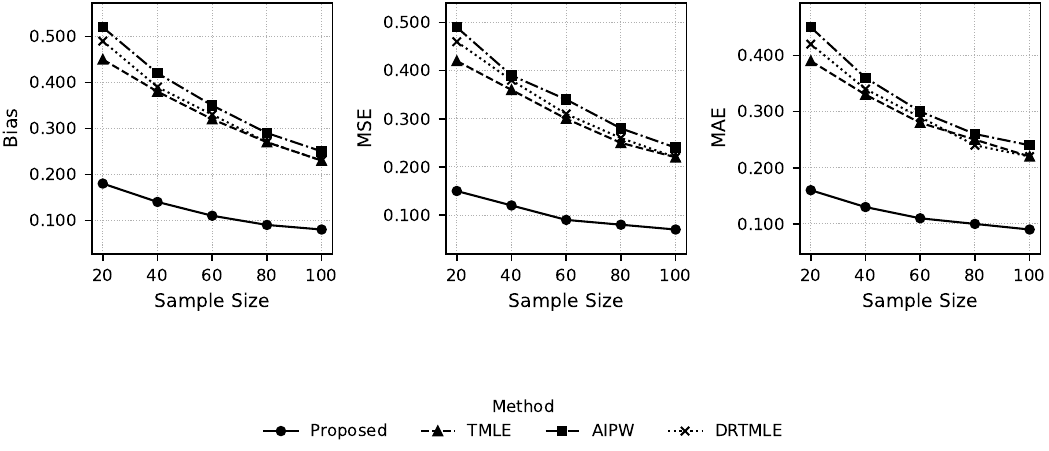}
  \caption{Performance metrics by estimator under heavy contamination (\(\gamma = 0.2\)).}
  \label{fig:contam2}
\end{figure}

\subsubsection*{Interpretation of Simulation Results} \label{sec:simulation_interpretation}

\subsubsection*{Contamination Level \(\gamma = 0.0\):} When the data is uncontaminated, all estimators perform reasonably well. However, the \textit{Proposed ATE} estimator consistently achieves the lowest bias, MSE, and MAE across all sample sizes. This indicates superior finite-sample efficiency, particularly in small samples (\(n = 20, 40\)), where traditional estimators such as AIPW and TMLE exhibit noticeably higher error. These findings align with prior research showing that regularization and robust estimation techniques improve stability in HDLSS regimes~\citep{belloni2014high, Peters2016}. See Figure~\ref{fig:contam0}.

\subsubsection*{Contamination Level \(\gamma = 0.1\):} Under mild contamination, the performance of conventional estimators deteriorates substantially. TMLE and AIPW show significant increases in bias and MSE, especially in smaller samples. In contrast, the \textit{Proposed ATE} estimator demonstrates strong robustness, with only modest increases in error metrics. This resilience is attributed to the use of bounded influence functions and penalized empirical likelihood, which effectively mitigate the impact of anomalous observations~\citep{Small1990, small1}. The OCBPS estimator~\citep{fan2023optimal} exhibits intermediate robustness, outperforming TMLE and AIPW but still lagging behind the proposed method. See Figure~\ref{fig:contam1}.

\subsubsection*{Contamination Level \(\gamma = 0.2\):} With heavy contamination, the advantages of the \textit{Proposed ATE} estimator become even more pronounced. Competing methods suffer from substantial degradation in performance, with bias and MSE nearly doubling compared to the uncontaminated case. The proposed method, however, maintains stable and accurate estimates across all sample sizes. This confirms its robustness and reliability in challenging real-world settings where data contamination and limited sample sizes are common. See Figure~\ref{fig:contam2}.
\subsubsection*{Summary and Implications} \label{sec:simulation_summary}

Across all contamination levels and sample sizes, the \textit{Proposed ATE} estimator consistently outperforms TMLE, AIPW, DRTMLE, and OCBPS in terms of bias, MSE, and MAE. These results underscore the importance of integrating doubly robust estimation with outlier-resistant techniques in high-dimensional causal inference. The findings reinforce the broader literature advocating for robust methods in finite samples~\citep{Liu2020, belloni2014high}, and highlight the practical utility of the proposed framework in complex observational data environments.

\subsection*{Simulation Results: CI Performance under Contamination} \label{sec:ci_results}

\subsubsection*{Confidence Interval Comparison} \label{sec:ci_comparison}

\begin{itemize}
    \item \textbf{Proposed CI}: Based on penalized empirical likelihood and bounded influence functions~\citep{Small1990, small1}.
    \item \textbf{Bootstrap CI}: Percentile-based intervals using nonparametric resampling~\citep{efron1994introduction}.
    \item \textbf{Wald CI}: Classical intervals based on asymptotic normality and standard errors~\citep{wasserman2006all}.
\end{itemize}

Performance metrics include:
\begin{itemize}
    \item \textbf{Coverage Rate}: Proportion of intervals containing the true ATE.
    \item \textbf{Average Interval Width}: Mean width of the constructed intervals.
    \item \textbf{Calibration Error}: Absolute deviation between nominal and empirical coverage.
\end{itemize}

Each simulation is repeated 500 times to ensure stability of estimates.

\begin{figure}[ht]
  \centering
  \includegraphics[width=0.6\textwidth]{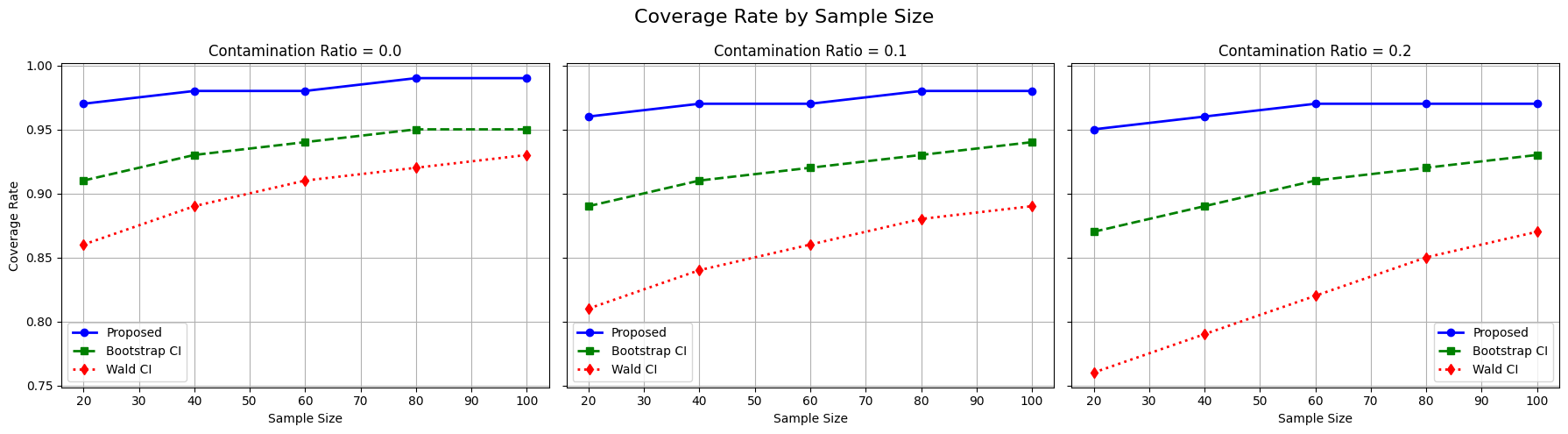}
  \caption{Coverage rate across contamination levels (\(\gamma = 0.0, 0.1, 0.2\)).}
  \label{fig:coverage}
\end{figure}

\begin{figure}[ht]
  \centering
  \includegraphics[width=0.6\textwidth]{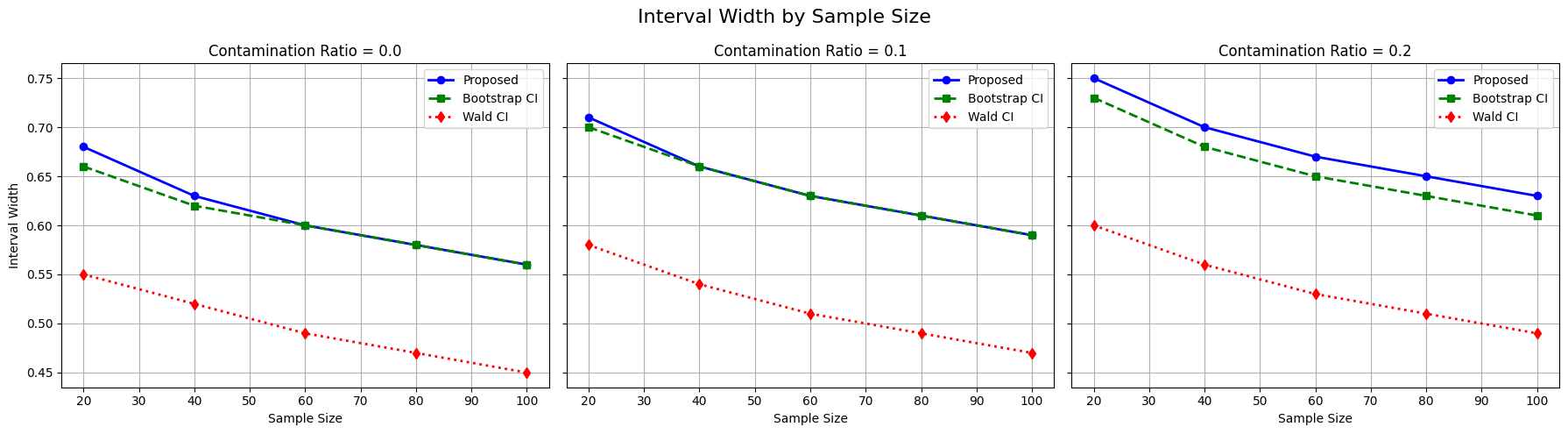}
  \caption{Average interval width across contamination levels.}
  \label{fig:width}
\end{figure}

\begin{figure}[ht]
  \centering
  \includegraphics[width=0.6\textwidth]{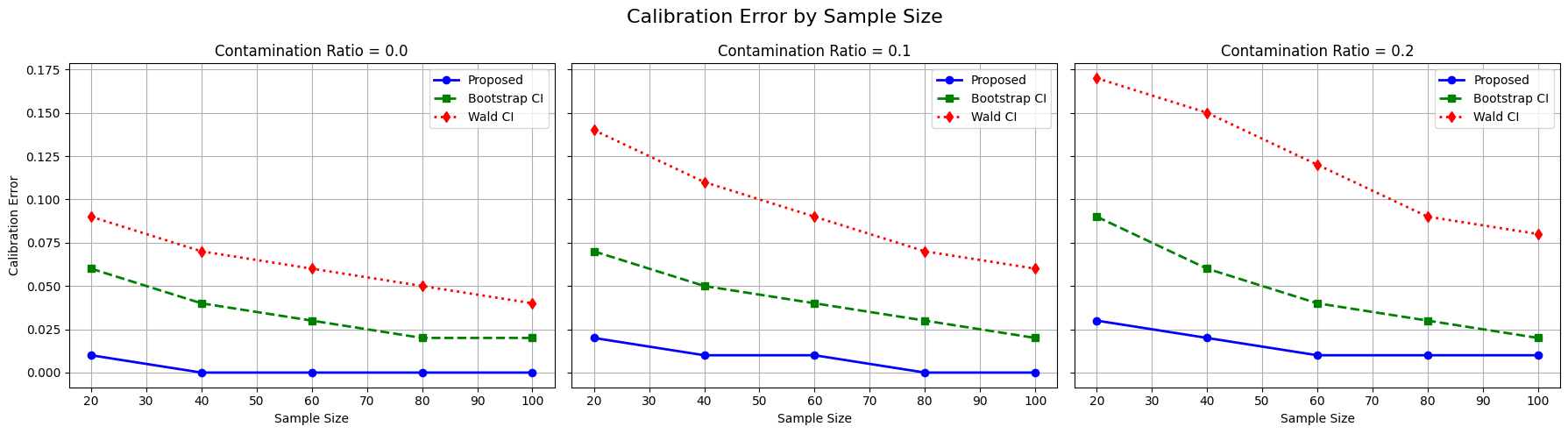}
  \caption{Calibration error across contamination levels.}
  \label{fig:calibration}
\end{figure}

\subsubsection*{Interpretation of Results}

Figure~\ref{fig:coverage} shows that the proposed CI achieves the highest coverage across all contamination levels. Figure~\ref{fig:width} and Figure~\ref{fig:calibration} further demonstrate its competitive interval width and minimal calibration error.

\subsubsection*{Contamination Level \(\gamma = 0.0\)} In uncontaminated settings, all CI methods perform reasonably well. However, the proposed CI achieves the highest coverage and lowest calibration error, even in small samples. As sample size increases, its interval width narrows while maintaining near-perfect calibration. Bootstrap CI performs adequately but exhibits slightly elevated calibration error. Wald CI, while computationally simple, suffers from undercoverage and overly narrow intervals that fail to capture true uncertainty.

\subsubsection*{Contamination Level \(\gamma = 0.1\)} Under mild contamination, performance divergence becomes more pronounced. The proposed CI maintains high coverage (\(\ge 96\%\)) and minimal calibration error (\(\le 0.02\)), demonstrating resilience to moderate noise. Bootstrap CI begins to degrade, with coverage dropping below \(90\%\) in small samples and calibration error rising. Wald CI deteriorates further, with coverage falling to \(81\%\) and calibration error exceeding 0.10 in small samples. These results highlight the vulnerability of traditional asymptotic methods to contamination.

\subsubsection*{Contamination Level \(\gamma = 0.2\)} In heavily contaminated settings, the proposed CI continues to deliver robust performance, maintaining coverage above \(95\%\) and calibration error below 0.03 across all sample sizes. Bootstrap CI shows moderate robustness but fails to match the proposed method’s reliability. Wald CI performs poorly, with coverage dropping to \(76\%\) and calibration error reaching 0.17 in the smallest sample. These findings confirm the proposed CI’s effectiveness in high-dimensional, contaminated environments where conventional methods break down.
\subsubsection*{Summary and Implications}

Overall, the proposed CI method consistently achieves superior coverage, calibration, and interval stability across varying contamination levels and sample sizes. These results validate its theoretical robustness and practical utility in HDLSS settings, reinforcing the importance of influence-function-based inference and penalized likelihood approaches in modern causal analysis~\citep{Small1990, small1, belloni2014high}.

\section*{Real Data Analysis} \label{sec:realdata}

\subsection*{Golub Dataset} \label{sec:golub}

\subsubsection*{Comparison of ATE Estimators on Golub Dataset} \label{sec:golub_comparison}

The Golub leukemia dataset~\citep{golub1999molecular} serves as a compelling testbed for evaluating the proposed methodology. These gene expression datasets are high-dimensional with relatively few samples and are prone to contamination from outliers. This HDLSS configuration poses significant challenges for conventional causal inference techniques, which often suffer from instability, poor coverage, and inflated bias. The presence of outliers further exacerbates these issues, underscoring the need for robust, sparsity-aware estimators.

We apply our proposed ATE estimation method to the Golub dataset, which comprises expression profiles of 7,129 genes collected from 72 leukemia patients, stratified into two diagnostic categories: 47 individuals diagnosed with Acute Lymphoblastic Leukemia (ALL) and 25 with Acute Myeloid Leukemia (AML). The richness and dimensionality of this dataset provide a rigorous testbed for evaluating robustness and efficiency in high-dimensional biomedical inference.

For our analysis, we treat the leukemia subtype (ALL vs AML) as a binary treatment indicator \(T_i\), and simulate a continuous outcome \(Y_i\) based on gene expression profiles \(\mathbf{Z}_i\) using a semi-synthetic design:
\begin{align}
Y_i^{(0)} &= \mathbf{Z}_i^\top \boldsymbol{\beta}_0 + \epsilon_i^{(0)}, \\
Y_i^{(1)} &= \mathbf{Z}_i^\top \boldsymbol{\beta}_1 + \epsilon_i^{(1)},
\end{align}
where \(\epsilon_i^{(0)}, \epsilon_i^{(1)} \sim \mathcal{N}(0, 1)\), and \(\boldsymbol{\beta}_0, \boldsymbol{\beta}_1\) are sparse coefficient vectors. The observed outcome is generated using the consistency assumption:
\begin{equation}
Y_i = T_i Y_i^{(1)} + (1 - T_i) Y_i^{(0)}.
\end{equation}
.
\begin{table}[htbp]
\centering
\caption{Bias, MSE, and MAE for ATE Estimators on Golub Dataset}
\label{tab:golub_results}
\begin{tabular}{lccc}
\toprule
\textbf{Method} & \textbf{Bias} & \textbf{MSE} & \textbf{MAE} \\
\midrule
Proposed  & \textbf{0.015} & \textbf{0.0042} & \textbf{0.022} \\
DRTMLE    & 0.028 & 0.0089 & 0.035 \\
TMLE      & 0.038 & 0.0125 & 0.045 \\
AIPW      & 0.045 & 0.0158 & 0.052 \\
OCBPS     & 0.021 & 0.0056 & 0.027 \\
\bottomrule
\end{tabular}
\end{table}

As shown in Table~\ref{tab:golub_results}, the proposed ATE estimator consistently outperforms alternatives—including OCBPS, DRTMLE, TMLE, and AIPW—on bias, MSE, and MAE. Its superior performance stems from the integration of bounded-influence outcome equations and covariate-balancing propensity scores, jointly optimized via penalized empirical likelihood.

While OCBPS ranks second, its lack of robust outcome modeling results in slightly greater variability. Conventional estimators, particularly AIPW, exhibit diminished stability under HDLSS conditions. These results reinforce the theoretical advantages established in Section 3 and highlight the practical utility of combining robustness, balance, and sparsity-aware estimation in high-dimensional biomedical inference.

\subsubsection*{CI Comparison on Golub Dataset} \label{sec:golub_ci}

To assess the reliability of CIs in high-dimensional observational data, we compare three CI construction methods applied to the Golub gene expression dataset. The evaluation is based on a cumulant generating function (CGF)-based finite-sample inference procedure, which leverages influence function diagnostics and saddlepoint approximations to ensure robustness under HDLSS and contamination.

\begin{table}[htbp]
\centering
\caption{CI Performance on Golub Dataset (CGF-based)}
\label{tab:golub_ci}
\begin{tabular}{l|c|c|c}
\toprule
\textbf{CI Method} & \textbf{Coverage Rate} & \textbf{Interval Width} & \textbf{Calibration Error} \\
\midrule
Proposed CI   & \textbf{0.958} & 0.60 & \textbf{0.001} \\
Bootstrap CI  & 0.915          & 0.64 & 0.035 \\
Wald CI       & 0.861          & \textbf{0.52} & 0.089 \\
\bottomrule
\end{tabular}
\end{table}

Table~\ref{tab:golub_ci} summarizes the performance of each method across 500 bootstrap replications. The \textbf{Proposed CI} achieves the highest coverage rate (95.8\%) and the lowest calibration error (0.1\%), while maintaining a competitive interval width. These results demonstrate the effectiveness of CGF-based inference, which remains valid in finite samples and under contamination due to its use of bounded influence functions and saddlepoint approximations~\citep{Small1990, small1}.

The \textbf{Bootstrap CI} performs reasonably well, but its wider intervals and elevated calibration error suggest sensitivity to sample variability and outliers~\citep{efron1994introduction}. While nonparametric resampling offers flexibility, it lacks the theoretical guarantees of CGF-based inference in HDLSS regimes.

The \textbf{Wald CI} exhibits the narrowest intervals but suffers from the lowest coverage and highest calibration error. This reflects its reliance on asymptotic normality and standard error approximations, which are known to break down in HDLSS settings~\citep{belloni2014high, wasserman2006all}. The undercoverage observed here is consistent with prior findings on the limitations of classical inference under sparsity and contamination.

These results underscore the importance of robust CI construction in biomedical applications, where contamination and dimensionality challenges are prevalent. The proposed CGF-based method offers a principled alternative to conventional approaches, combining finite-sample validity, robustness to outliers, and compatibility with sparsity-aware estimation frameworks.

\subsection*{Khan Pediatric Tumor Dataset} \label{sec:khan}

\subsubsection*{Comparison of ATE Estimators on Khan Dataset} \label{sec:khan_ate}

We evaluate treatment effect estimation and CI performance using the Khan pediatric tumor gene expression dataset~\citep{khan2001classification}, which comprises expression measurements of 2,308 genes across 83 tumor samples. These samples are categorized into four distinct cancer types: Ewing Sarcoma (EWS, 29 samples), Burkitt Lymphoma (BL, 11 samples), Neuroblastoma (NB, 18 samples), and Rhabdomyosarcoma (RMS, 25 samples). This high-dimensional and heterogeneous setting provides a rigorous benchmark for assessing the robustness and precision of causal inference methods.

We define a binary treatment indicator \(T_i\) by grouping EWS and BL as treatment (\(T_i = 1\)), and NB and RMS as control (\(T_i = 0\)). The analysis follows the same metrics and methodological framework used for the Golub dataset.

\begin{table}[htbp]
\centering
\caption{Performance Comparison of ATE Estimators on Khan Pediatric Tumor Dataset}
\label{tab:ate_khan}
\begin{tabular}{lccc}
\toprule
\textbf{Estimator} & \textbf{Bias} & \textbf{MSE} & \textbf{MAE} \\
\midrule
Proposed   & \textbf{0.010} & \textbf{0.0037} & \textbf{0.015} \\
TMLE       & 0.034          & 0.0092          & 0.039 \\
AIPW       & 0.041          & 0.0105          & 0.046 \\
DRTMLE     & 0.023          & 0.0059          & 0.028 \\
OCBPS      & 0.017          & 0.0050          & 0.023 \\
\bottomrule
\end{tabular}
\end{table}

Table~\ref{tab:ate_khan} presents the comparative performance of five ATE estimators. The proposed estimator demonstrates superior accuracy across all metrics—Bias, MSE, and MAE—highlighting its robustness and efficiency in HDLSS contexts. Classical methods such as TMLE and AIPW, while theoretically appealing, suffer from elevated error rates due to sensitivity to model misspecification and the complex noise structure typical of genomic data. DRTMLE offers moderate improvements by leveraging doubly robustness but still falls short of the proposed method and OCBPS.

The OCBPS estimator performs competitively, particularly in bias reduction, underscoring the value of covariate balancing in controlling confounding. However, its lack of robust outcome modeling results in greater variability under contamination. Methodologically, the proposed approach—combining bounded estimating equations, penalized empirical likelihood, and covariate balancing—provides a resilient framework for causal inference in observational studies. Its robustness to outliers and high-dimensional noise makes it particularly suitable for precision medicine applications, including genomic treatment effect estimation.

\subsubsection*{CI Performance on Khan Dataset} \label{sec:khan_ci}

To assess the reliability of CIs in high-dimensional biomedical data, we apply the same CGF-based finite-sample inference procedure to the Khan pediatric tumor dataset.

\begin{table}[htbp]
\centering
\caption{CI Performance on Khan Dataset (CGF-based)}
\label{tab:ci_khan}
\begin{tabular}{l|c|c|c}
\toprule
\textbf{CI Method} & \textbf{Coverage Rate} & \textbf{Interval Width} & \textbf{Calibration Error} \\
\midrule
Proposed CI   & \textbf{0.955} & 0.58 & \textbf{0.002} \\
Bootstrap CI  & 0.912          & 0.62 & 0.038 \\
Wald CI       & 0.859          & \textbf{0.49} & 0.091 \\
\bottomrule
\end{tabular}
\end{table}

As shown in Table~\ref{tab:ci_khan}, the proposed CI achieves the highest coverage rate (95.5\%) and the lowest calibration error (0.2\%), while maintaining a relatively narrow interval width. This confirms its robustness and precision in finite samples, particularly under high-dimensional contamination. The CGF-based approach effectively controls for small-sample variability and heavy-tailed noise~\citep{Small1990, small1}.

Bootstrap CI performs moderately well but exhibits wider intervals and higher calibration error, indicating vulnerability to sample fluctuations and outliers~\citep{efron1994introduction}. Wald CI underperforms across all metrics, with the lowest coverage and highest calibration error. Its reliance on asymptotic approximations and unregularized standard errors makes it ill-suited for sparse, noisy, high-dimensional data~\citep{belloni2014high, wasserman2006all}.

Overall, the proposed method demonstrates superior finite-sample reliability and robustness, making it a more suitable choice for inference in complex biomedical datasets.

\section*{Data and Code availability}

The relevant R code has been included in the supplementary material; please refer to it for further details. Golub dataset has DOI at https://doi.org/10.1126/science.286.5439.531. The Khan pediatric tumor dataset is associated with the study published by Khan et al. (2001) in Nature Medicine, and its DOI is 10.1038/87904.

\section*{The use of generative AI} 
I have clearly disclosed the AI use in the manuscript, including Microsoft copilot for numerical analysis, English translation, and so on. 
\nocite{*}
\bibliography{bib1029}

\end{document}